# Thallium under extreme compression

C. Cazorla[1,2], S. G. MacLeod[3,4,5], D. Errandonea[6], K. A. Munro[7], M. I. McMahon[5,7], and C. Popescu[8]

[1]School of Materials Science and Engineering, UNSW Australia, Sydney NSW 2052, Australia

[2]Integrated Materials Design Centre, UNSW Australia, Sydney NSW 2052, Australia

[3]Atomic Weapons Establishment, Aldermaston, Reading, RG7 4PR, United Kingdom

[4]Institute of Shock Physics, Imperial College London, SW7 2AZ, United Kingdom

[5]Research Complex at Harwell, Didcot, Oxon, OX11 0FA, United Kingdom

[6]Departamento de Física Aplicada-Instituto de Ciencia de Materiales, Matter at High Pressure (MALTA) Consolider Team, Universidad de Valencia, Edificio de Investigación, C/Dr. Moliner 50, Burjassot, 46100 Valencia, Spain

[7]SUPA, School of Physics and Astronomy, and Centre for Science at Extreme Conditions, The University of Edinburgh, Edinburgh, EH9 3FD, United Kingdom

[8]CELLS-ALBA Synchrotron Light Facility, 08290 Cerdanyola, Barcelona, Spain

**Abstract:** We present a combined theoretical and experimental study of the high-pressure behavior of thallium. X-ray diffraction experiments have been carried out at room temperature up to 125 GPa using diamond-anvil cells, nearly doubling the pressure range of previous experiments. We have confirmed the hcp-fcc transition at 3.5 GPa and determined that the fcc structure remains stable up to the highest pressure attained in the experiments. In addition, HP-HT experiments have been performed up to 8 GPa and 700 K by using a combination of x-ray diffraction and a resistively heated diamond-anvil cell. Information on the phase boundaries is obtained, as well as crystallographic information on the HT bcc phase. The equation of state for different phases is reported. *Ab initio* calculations have also been carried out considering several potential high-pressure structures. They are consistent with the experimental results and predict that, among the structures considered in the calculations, the fcc structure of thallium is stable up to 4.3 TPa. Calculations also predict the post-fcc phase to have a close-packed orthorhombic structure above 4.3 TPa.

# 1. Introduction

Studying materials at extreme pressures provides insights into the deep interiors of large planets and chemistry under extreme conditions. Many metals have been studied by x-ray diffraction (XRD) up to 2 - 3 megabars (Mbar) to determine structural phase transitions induced by pressure [1 - 6]. Theoretical studies have extended the pressure range of study to predict stable structures of metals at multi-terapascal (TPa) pressures [7, 8]. Systematic investigations of the high-pressure behavior of metals have recently concentrated on alkaline earths [1], transition metals [2], and lanthanides [9, 10]. By contrast, elemental post-transition metals like thallium (Tl) have been poorly studied. This is probably due to its high toxicity and tendency to oxidize and chemically react. Tl is important from a technological point of view both as a pure element or when forming alloys or compounds [11]. In particular, when alloyed with halide elements, wide band-gap materials are formed with promising spectroscopy applications [11].

At ambient conditions Tl crystallizes in the hexagonal close-packed (hcp) structure [12]. The first experiments under compression on this element date from half a century ago [13 – 15]. In particular, its high pressure (HP) structural sequence has been studied using angle-dispersive XRD (ADXRD) and energy-dispersive XRD (EDXRD), at room temperature (RT) up to 68 GPa [12, 16, 17]. A phase transition to the face-centered cubic (fcc) structure occurs at 3.8 GPa. At HP and high-temperature (HT), Tl transforms to the bcc structure before melting [18]. After nearly two decades since the last HP study on Tl was carried out, we consider it timely to study again the structural properties of Tl under compression using current state-of-the-art techniques. This is due, in particular, to the relatively low maximum compression of previous studies, the discovery of complex structural forms [19] in other elemental metals, and the prediction of Miao and Hoffmann [20] that Tl will form electride structures at extreme compression. We report here,

ADXRD measurements carried out at RT using a diamond-anvil cell (DAC) up to nearly 125 GPa. We report also HP-HT measurements carried out in a resistively heated DAC up to 700 K and 8 GPa. Finally, we report *ab initio* calculations up to 7.5 TPa. Experiments and calculations agree well up to 125 GPa, which improves confidence in our theoretical predictions at higher pressure. In particular, calculations predict that a phase transition from fcc to an orthorhombic phase occurs beyond 4.3 TPa. The equation of state (EOS) of the hcp, fcc, and body-centered cubic (bcc) phases will be also reported

2. **Experimental details**

We performed two experiments at RT and three experiments at HT. Several pressure cells were used in the different experiments performed. For the RT compression experiments to 125 GPa, gas-membrane driven DACs, equipped with beveled diamonds (culet diameters 150 μm and 300 μm) and a rhenium (Re) gasket, were loaded with Tl powder of 99.99% purity purchased from Aldrich Chemicals. We decided not to include a pressure-transmitting medium (PTM) in our experiments in order to reduce the possibility of oxidation of thallium during DAC loading and the exposure of the scientist to a hazardous chemical such as thallium. This approach has been proved to satisfactorily work with metals that possess a small bulk modulus like Tl, e.g magnesium (Mg) [1]. Powder ADXRD data were collected on compression using the MSPD-BL04 beamline at the ALBA synchrotron, using an x-ray wavelength of 0.4246 Å and a beam size of 5 μm x 5 μm. One RT experiment was carried out using gold (Au) [21] as the pressure marker and the other using tantalum (Ta) [22]. All DACs (RT and also HT experiments) were loaded with small pieces of Tl with fresh surfaces in an inert atmosphere filled glove box in order to preserve the chemical integrity of the samples. Data were collected to 125 GPa. Unfortunately, at this pressure, the anvils failed, terminating the experiment.

For the HT resistive-heating studies we used three gas-membrane driven DACs equipped with diamonds with either 250 or 300 µm culets and loaded with same Tl in the same manner as described above. Either Au foil or Ta powder were used as the pressure marker [23]. No PTM was used for the reasons described in the RT experiments. The DACs were contained within a custom-built vacuum vessel designed for HP-HT experiments [1, 24]. The DACs were heated using Watlow 240 V (rated at 4.65 W/cm$^2$) coiled heaters wrapped around the outside of the DACs. The temperature was measured using a K-type thermocouple attached to one of the diamond anvils, close to the gasket. The accuracy of the thermocouple on the temperature range covered by the experiments is 0.4%. The diffraction data were collected using the same beamline and x-ray wavelength employed for the RT experiments.

In all the experiments, the two-dimensional diffraction patterns were collected on a Rayonix CCD and integrated azimuthally using FIT2D [25]. The resulting one-dimensional diffraction profiles were analyzed by Le Bail fitting. Since no PTM was utilized in our DACs, the samples may have experienced non hydrostatic-pressure conditions, leading to a small inaccuracy in the sample pressure. Tl is a very soft metal (Mohs hardness 1.2 MPa), and so we anticipated any such effects to be non-significant. Indeed as we previously observed for Mg [1], we found that for Tl the effects of non-hydrostatic-pressure conditions are negligible below 50 GPa. For higher pressures, they were taken into account following the analysis of Singh and Kenichi [26, 27]. In the Ta pressure marker (used for all measurements above 12.4 GPa), beyond 50 GPa, there was a small systematic overestimate of the unit-cell volume, which increases upon compression, leading to a maximum pressure underestimate of 1 GPa at 125 GPa.

3. **Computational methods**

We used the generalised gradient approximation to DFT devised by Perdew,

Burke, and Ernzerhof (GGA-PBE) [28], and implemented in the VASP package [29]. We employed the projector-augmented wave method [30] to represent the ionic Tl cores and considered the following electrons as valence states: $5d^{10}$, $6s^2$, and $6p^1$. Relativistic spin-orbit effects were taken into account in our calculations. Wave functions were represented in a plane-wave basis truncated at 600 eV. For integrations within the Brillouin zone (BZ), we employed dense Γ-centered **k**-point grids of 16 x 16 x 12 in the hcp structure, and equivalent ones in the rest of the structures. Geometry relaxations were performed using a conjugate-gradient algorithm that changed the volume and shape of the unit cell and imposed a tolerance on the atomic forces of 0.01 eV·Å$^{-1}$. Using these parameters we obtained enthalpy energies that were converged to within 0.5 meV per atom.

We calculated the vibrational phonon frequencies of a series of different crystal structures by using the small-displacement method [31, 32] and the PHON package due to Alfè [33]. The quantities with respect to which our phonon calculations need to be converged are the size of the supercell, the size of the atomic displacements, and the numerical accuracy in the calculation of the atomic forces and BZ sampling [31, 32]. We found the following settings to fulfil our accuracy requirements (i.e., quasi-harmonic free energies converged to within 5 meV/atom) in the hexagonal hcp structure: 5 x 5 x 4 supercells (containing 200 atoms), atomic displacements of 0.02 Å, and Γ-centered **k**-point grids of 3 x 3 x 3 for BZ sampling. Regarding the calculation of the atomic forces using VASP, we computed the non-local parts of the pseudopotential contributions in reciprocal, rather than real, space. Equivalent technical parameters were adopted in the rest of the phonon calculations, adapting the density of k-points to each crystal structure. In using this PHON code. We exploited the translational invariance of the system to impose the three acoustic branches to be exactly zero at the Γ **q**-point, and used central differences in the atomic forces (i.e., we considered positive and negative atomic displacements). Finally, the electron localization function (ELF) and electronic density of states calculations were also performed with VASP.

## 4. Results

### A. Room temperature XRD experiments

We have extended the pressure range of previous XRD experiments from 68 to 125 GPa. Previously the HP crystal structure of Tl was studied using ADXRD and EDXRD, but there was a significant amount of scatter in the hcp data [12, 16, 17]. Here we report high-resolution ADXRD measurements. Fig. 1 shows a selection of XRD patterns. We found that the low-pressure phase (hcp structure) remains stable up to 3.4 GPa. At 3.5 GPa we found evidence of a phase transition to the fcc structure. This result is in good agreement with the results reported in the most recent study by Schulte *et al.* [17]. Upon further compression at RT we did not find evidence of any other phase transition. The ADXRD patterns can be assigned to the fcc structure up to the highest pressure of 125 GPa obtained in our experiments. In one of the experiments we observed the coexistence of the hcp and fcc phases from 2.8 to 5 GPa. The observed hcp-fcc transition is reversible with a hysteresis of less than 2 GPa. Consequently only the hcp phase is recovered after full decompression to ambient pressure.

From two independent ADXRD experiments we have determined the pressure dependence of the unit-cell parameters of hcp Tl. The results are summarized in Fig. 2. In the figure the experimental results are compared with the results of our *ab initio* calculations. The agreement between experiments and calculations is very good. We found that the compression of the hcp structure is anisotropic. In particular, the *c*-axis is less compressible than the *a*-axis. This can be clearly seen in the inset of Fig. 2, where we represent the *c/a* ratio versus pressure. This ratio increases from 1.596(3) at ambient pressure to 1.610(3) at 3.4 GPa. In Fig. 2 we also represent the pressure dependence of the unit-cell volume. We found that both the experimental and theoretical results can be described by a 2$^{nd}$ order Birch-Murnaghan equation of state (BM EOS) [34]. From the experiments we determined the following EOS parameters: volume at ambient pressure (we used the volume/atom to facilitate comparison among structures), $V_0 = 28.7(1)$ Å$^3$ and bulk modulus $B_0 = 31(2)$ GPa; with its pressure derivative fixed to $B_0' = 4$. From the calculations we obtained: $V_0 = 29.36$ Å$^3$ and bulk modulus $B_0 = 38.9$ GPa ($B_0' = 4$). The agreement between both sets of parameters is reasonable. They also compare well with previous results reported in the literature [12, 16, 17].

In Fig. 3 we represent the pressure dependence of the unit-cell volume of the fcc phase as compared with the volume of the hcp phase. In the figure (top panel), it can be seen that the scatter of data present in previous experiments is not observed in our experiments, from which we obtained a smooth pressure dependence for the lattice parameters and volume. This improvement of the quality of the experimental results allows us to observe at the hcp-fcc phase transition a small discontinuity of the unit-cell volume (see center panel of Fig. 3). The determined volume collapse ($\Delta V/V_0$) is -0.5%, which is larger than the uncertainty of the volume determination from our experiments. A similar discontinuity (-0.8%) has been determined from our DFT calculations, which slightly overestimate the unit-cell volume of both phases. The detected volume discontinuity suggests that the hcp-fcc phase transition is a first-order transition. On the other hand, the observed phase coexistence suggests a transformation mechanism in which some kind of rearrangement or shuffling of the lattice points gradually transforms the hcp structure into an fcc structure [35]. In the bottom panel of Fig. 3 it can be seen that experiments and calculations give a similar evolution of the volume with pressure in the pressure range covered by our XRD measurements. From the experiments, for the HP fcc structure, considering the whole pressure range covered by the experiments, we determined the following EOS parameters by assuming a 2$^{nd}$ order BM EOS: $V_0$ = 28.2(2) Å$^3$ and $B_0$ = 48(6) GPa ($B_0'$ = 4). If a 3$^{rd}$ order BM EOS is used we got $V_0$ = 28.2(2) Å$^3$, $B_0$ = 50(5) GPa, and $B_0'$ = 3.2(3). Therefore, the fcc phase is apparently not only more dense than the hcp phase but also slightly less compressible. The agreement between our experimental and theoretical results is good; with DFT providing $V_0$ = 28.84 Å$^3$, $B_0$ = 44.78 GPa, and $B_0'$ = 3.94. This agreement confirms that the effects of non-hydrostatic pressure conditions (induced by the fact that no pressure medium was used in the experiments) on the compressibility of Tl are small.

### B. High temperature XRD experiments

We have explored the phase diagram of Tl at HT up to 8 GPa. The highest temperature we reached was 700 K. We have observed only the hcp, fcc, and bcc phases, as expected from the phase-diagram proposed by Tonkov [18]. In the present study we obtained *in situ* information on the crystal structure of these three phases. Fig. 4 shows three selected HT-HP XRD patterns to illustrate the identification of the different crystal structures. Fig. 5 shows the phase diagram proposed by Tonkov and the P-T points where we have observed the different phases. The hcp-bcc-fcc triple point is located at 3.75 GPa and 400 K. An interesting observation was that in three cases we reached a temperature corresponding to the melting temperature of Tl (solid circles in the figure). In all three cases we observed a fast and uncontrollable chemical reaction between Tl and the Re gasket. The liquid Tl dissolved the Re gasket and caused it to collapse. This discovery suggests that liquid Tl is much more reactive than solid Tl. In addition, we also observed that in the bcc phase Tl tended to recrystallize and form single crystals.

By comparing the experiments carried out at 390 – 400 K with those carried out at RT, we determined that the *c/a* is not affected by temperature, within the accuracy of the experiments. In addition, we determined for the hcp-fcc transition, that the volume collapse at 390 – 400 K is the same as determined at RT (-0.5 %). At this temperature, we observed the hcp phase at 3.4 GPa and then the appearance of the fcc phase in a subsequent pressure increase to 3.7 GPa. This observation is consistent with the representation of the hcp-fcc boundary as a nearly vertical line. For the bcc phase, we determined that at 1.5 GPa and 467 K the unit-cell parameter is 3.882(5) Å. From the results measured at 467 K we also estimated the bulk modulus of the bcc phase using a $2^{nd}$ order BM EOS ($B_0' = 4$). We obtained $B_0 = 42(4)$. This value is similar to the value that can be calculated from the elastic constants reported by Iizumi [36]. Consequently, the bcc phase has a bulk modulus slightly smaller than the fcc phase (in which $B_0$ is reduced from 48(6) at 300 K to 45(6) at 467 K); however both values nearly agree within error bars. For the hcp-bcc transition we determined at 450 K that the transition occurs together with a volume collapse of -0.3 %. Furthermore, at 467 K there is a volume collapse of -0.3% when going from the bcc to the fcc phase. Therefore, we can conclude that the relative isothermal volumes are: $V_{fcc} < V_{bcc} < V_{hcp}$. On the other hand, according to our calculations, the entropy (S) of the bcc is larger than that of the hcp and fcc phases (which have a very similar entropy). Thus, the fact that at the hcp-bcc transition $\Delta V < 0$ and $\Delta S > 0$, explains why the hcp-bcc boundary has a negative slope (dT/dP = -26 K/GPa), since according to the Clausius–Clapeyron relation $\frac{dT}{dP} = \frac{\Delta V}{\Delta S}$. In the case of the bcc-fcc transition we have $\Delta V < 0$ and $\Delta S < 0$, which is consistent with the positive slope of the bcc-fcc phase boundary. For this boundary, at P < 5 GPa, dT/dP is approximately 120 K/GPa; however at P > 5 GPa the phase boundary bends towards the pressure axis becoming dT/dP = 50 K/GPa. This slope it quite similar to the melting slope, as can be seen in Fig. 5 where above 5 GPa the bcc-fcc phase boundary is nearly parallel to the melting curve. Therefore, the present results on the HP-HT phase diagram suggest that a fcc-bcc-liquid triple point should not be expected at higher pressure if the melting slope is not modified by compression. Finally, the DFT prediction that ΔS ≈ 0 at the hcp-fcc phase boundary is consistent with an essentially vertical hcp-fcc phase boundary that has been determined from experiments.

### C. Calculations

#### C.1 Gigapascal regime

In order to study the structural stability of Tl theoretically, besides the three crystal phases that are observed in the experiments, namely hcp [space group (s.g.) 194], bcc (s.g. 229), and fcc (s.g. 225), we analysed some other phases within the pressure interval 0 < P < 150 GPa, aiming at searching for possible solid-solid phase transitions. In particular, we calculated the zero-temperature enthalpy (H = E + P·V) of the following crystal structures: simple cubic (s.g. 221), simple hexagonal (s.g. 194), double hcp (dhcp, s.g. 194), omega (s.g. 191), distorted fcc (dfcc, s.g. 166), post-dfcc (s.g. 12), and a monoclinic structure described with space group *C2/m* (s.g. 12 and which we will name simply as *C2/m*) and which is related to the α'' structure of cerium described by McMahon and Nelmes [37]. These structures were selected by applying the method of data mining considering the crystal structures known from highly compressed lanthanide [38], alkali [39], alkaline earth [40], and other elemental metals [41, 42]. We note that although the post-dfcc and *C2/m* phases are both monoclinic and share the same space group, they are different structures. The first one (post-dfcc) is the structure found in gadolinium around 65 GPa [38]. The other comes from the application of random displacements in the atoms of the former structure and a subsequent *ab initio* structure relaxation. Actually, we found that the post-dfcc phase is transformed into the fcc by effect of pressure, in contrast to the *C2/m* structure.

Fig. 6 shows the zero-temperature enthalpy per atom of the analysed crystal structures relative to the ground-state hcp phase and expressed as a function of pressure. Results obtained for some of the investigated structures are not shown therein because the corresponding energies turned out to be too large in comparison to the ground-state phase. At P = 0 GPa, only the hexagonal omega, hexagonal double hcp, cubic fcc, cubic bcc, and monoclinic *C2/m* were found to be energetically competitive with respect to the hexagonal hcp phase. Upon compression, however, the hexagonal omega phase also turned out to be disfavoured. At T = 0 K and a pressure of 8.8 GPa, in agreement with the experiments, we found that Tl transforms from the hexagonal hcp to the cubic fcc structure. However, the transition pressure is slightly overestimated. The fcc remains the most stable structure up to the TPa regime (see next section). As shown in Figs. 7 and 8, both crystal structures (hcp and fcc) are vibrationally stable over the pressure range relevant to the hcp → fcc transformation.

Fig. 9 shows the vibrational phonon frequencies calculated in the cubic bcc structure at several pressures. We found that at low-temperature, within the pressure interval 0 < P < 8 GPa, this phase presents some phonon instabilities (i.e., imaginary phonon frequencies represented with negative values) at the reciprocal **q** vectors (0,½,0) and (½, ½, ½). This result indicates that Tl in the cubic bcc structure highly anharmonic. We found also that near equilibrium the electronic entropy in the bcc phase is larger than that in the hcp and bcc structures in the temperature interval 0 < T < 1000K; we note, however, that the electronic entropy differences between the bcc and rest of energetically competitive structures amount to less than 1 meV/atom, which is much smaller than the corresponding enthalpy energy differences of ~10 meV/atom. In fact, these theoretical findings are consistent with the experimental observations: at moderate pressure, the cubic bcc phase is entropically stabilised over the hexagonal hcp and cubic fcc phases by effect of temperature.

At a pressure higher than 175 GPa and zero temperature, we found that the monoclinic *C2/m* structure is reduced into an orthorhombic *Cmcm* structure (s.g. 63). In particular the monoclinic α angle of the *C2/m* structure gradually changes upon compression reaching a 90° value at ~ 200 GPa. Beyond this pressure, α remains equal to 90º, and the crystal structure can be described with the orthorhombic *Cmcm* space group. This structure becomes the most stable structure in the TPa regime. The orthorhombic *Cmcm* structure is close-packed with a packing factor equal to that of the fcc and hcp structures. The structural transformation from the *C2/m* structure into the *Cmcm* space corresponds to a second-order phase transition since the associated volume change is null and the two involved structures fulfil a group-subgroup relationship (*C2/m* ⊃ *Cmcm*) [43].

### C. 2   Terapascal regime

We extended our analysis of the relative stability between different crystal structures in Tl up to the TPa regime (i.e., pressures of the order of $10^3$ GPa). We found that the cubic fcc structure remained the ground state up to a pressure of 4.33 TPa. At this point, the orthorhombic *Cmcm* phase became the new phase of minimum enthalpy (see Fig. 10). This structure is fully derived from our theoretical simulations and to the best of our knowledge it has never been observed before in elemental metals. The *Cmcm* structure can be seen as a distorted fcc with alternative layers sliding along [010]. The existence of such a structure could be experimentally confirmed in the near future since TPa pressures may be achievable thanks to the development of micro-ball nano-diamond anvils [44]. At 5.3 TPa, the *Cmcm* structure has unit-cell parameters *a* = 1.9223 Å, *b* = 3.3367 Å, and *c* = 3.1182 Å. In this structure all the atoms are at the Wyckoff position 4c (0, y, 0.25), in which only the coordinate y = 0.3339 is not determined by symmetry. According to our calculations, the volume change ($\Delta V/V_0$) associated to this fcc-*Cmcm* transformation is -0.4 % with $V_{fcc}$ = 5.14 Å$^3$/atom. Interestingly, near the transition at T = 0 K the hexagonal hcp phase becomes again energetically more favourable than the cubic fcc. This finding indicates that a re-entrant phase behaviour in ultra-compressed Tl could be possible at T > 0 K.

Our phonon calculations performed in the TPa regime show that the three most energetically competitive phases in Tl, namely the hcp, fcc, and *Cmcm* structures, are all vibrationally stable (see Fig. 11). We also computed the density of electronic states in these three phases in order to check whether a metal → insulator could be induced by the effect of pressure. Our first-principles results show, however, that ultra-compressed Tl remains metallic in any of the analysed structures (see Fig. 12).

Before concluding we would like to comment on the prediction that Tl becomes an electride under extreme compression [20]. In order to explore this possibility, we have calculated the electron localization function in the orthorhombic *Cmcm* and cubic fcc phases of Tl at a pressure of 7.5 TPa. Our results show that Tl is not an electride (in contrast to Miao and Hoffman's hypothesis). This can be seen in Fig. 13, where we show the electron localization function isosurface maps calculated in both structures. In both of them, the highest density of electrons is found around the nuclear centres and in the bonds created between them, no significant accumulation of electrons is found in the interstitials. The main reason for this is the well-known *sp*→*d* electron transfer effect that is induced by pressure [45]. As compression is raised, the energy of the valence *s* and *p* orbitals increases more rapidly than that of the conduction *d* orbitals. Eventually, the conduction *d* orbitals become valence and electrons are transferred to them from the *s* and *p* orbitals. This can be seen clearly in Fig. 14 by comparing the projected electronic density of states (pDOS). As pressure is increased, the population of *d* orbitals around the Fermi energy level increases. The same thing occurs in the fcc phase. The model that was presented in Miao and Hoffman's [20] did not take into consideration this pressure-induced $sp \rightarrow d$ electron transfer effect. This is the main reason why their prediction that Tl should become an electride under extreme compression is not correct.

## V. Concluding remarks

The high-pressure structural behavior of Tl has been explored by combining ADXRD measurements and *ab initio* calculations. Experiments and calculations agree well up to 125 GPa, the maximum pressure achieved in the experiments. The transition from hcp to fcc is confirmed to occur at 3.5 GPa. The fcc phase remains stable up to 125 GPa. We also report HP-HT experiments. From our studies information on the crystal structure of different phases of Tl is obtained. This information allowed us to better understand the HP-HT phase diagram of Tl and to determine an accurate EOS for the hcp, bcc, and fcc phases of Tl. In particular we determined the volume discontinuities observed at the hcp-bcc, hcp-fcc, and bcc-fcc transitions. The extension of the calculations to the terapascal regime predicts that a phase transition from fcc to an orthorhombic phase occurs beyond 4.3 TPa. Finally, our calculations showed that under extreme compression Tl does not become an electride, ruling out a previous hypothesis. We hope our results will trigger further studies of Tl which up-to-now have remained unexplored for nearly two decades.

**Acknowledgements**

CC acknowledges support from the Australian Research Council under the Future Fellowship funding scheme (project number FT140100135). Computational resources were provided by the Australian Government through Magnus under the National Computational Merit Allocation Scheme. Part of the research was supported by the Spanish government MINECO under Grants No: MAT2013-46649-C4-1-P and MAT2015-71070-REDC. ©British Crown Owned Copyright 2016/AWE. Published with permission of the Controller of Her Britannic Majesty's Stationary Office. MIM is grateful to AWE for the award of a William Penney Fellowship. Experiments were performed at MSPD beamline at ALBA Synchrotron Light Facility with the collaboration of ALBA staff.

**Figure captions**

**Figure 1:** Selected RT XRD patterns measured at different pressures. Pressures are indicated in the figure. Ticks identifying the Bragg peaks of the different phases of Tl and the pressure markers (Au or Ta) are shown.

**Figure 2:** Lattice parameters and volume versus pressure for the hcp phase. Circles: data obtained using Ta as pressure marker. Squares: data obtained using Au as pressure marker. The inset shows the *c/a* ratio. In the top panel, solid lines are the results of *ab initio* calculations. In the bottom panel, the solid lines show the theoretical results and the dashed line the EOS described in the text.

**Figure 3:** (color online) Volume as a function of pressure for the hcp and fcc phases. Top: Present results are red solid circles (hcp Ta marker), red empty circles (fcc Ta marker), red solid squares (hcp Au marker), and red empty squares (fcc Au marker). Previous results are from Ref. 12 (stars), Ref. 16 (diamonds), and Ref. 17 (triangles) corresponding the solid symbols to the hcp phase and the empty symbols to the fcc phase. Center: present results compared with calculations up to 20 GPa to facilitate the identification of the volume discontinuity at the hcp-fcc transition. Solid lines are the EOS determined from the experiments (see text) and dashed lines represent the theoretical results. Bottom: same but extending the pressure range up to 125 GPa.

**Figure 4:** Representative XRD patterns measured at HP-HT illustrating the identification of the three phases experimentally found. The pressure, temperature, and crystal structure are indicated in the figure.

**Figure 5:** P-T phase diagram of Tl. The solid lines represent the phase diagram of Tonkov [18]. The solid squares are the P-T conditions were the hcp phase is observed. The empty squares the P-T conditions were the fcc phase is observed. The empty circles the P-T conditions were the bcc phase is observed. There are symbols which are half

solid representing the observation of phase coexistence. The solid circles indicate the P-T conditions were a rapid chemical reaction between Tl and Re is detected (causing the collapse of the Re gasket). These P-T conditions agree with the melting curve of Tonkov [18].

**Figure 6:** (color online) Zero-temperature enthalpy energy of several crystal structures expressed as a function of pressure and referred to the ground-state hcp structure.

**Figure 7:** (color online) Vibrational phonon frequencies of the hexagonal hcp structure at several pressure points.

**Figure 8:** (color online) Vibrational phonon frequencies of the cubic fcc structure at several pressure points.

**Figure 9:** (color online) Vibrational phonon frequencies of the cubic bcc structure at several pressure points.

**Figure 10:** (color online) Zero-temperature enthalpy energy of several crystal structures expressed as a function of pressure and referred to the hexagonal hcp structure.

**Figure 11:** (color online) Vibrational phonon frequencies of the orthorhombic *Cmcm* structure.

**Figure 12:** (color online) Density of electronic states around the Fermi energy level for several energetically competitive structures at P ~ 5000 GPa.

**Figure 13:** Electron localization function (ELF) in the orthorhombic *Cmcm* and cubic fcc phases of Tl at 7.5 TPa. An isosurface of value 2.85 (arb. unit) is represented in yellow. Regions in which the density of electrons is high (low) are represented in blue (red). No significant accumulation of charge is observed in the interstitials.

**Figure 14:** Partial density of electronic states calculated in the orthorhombic *Cmcm* phase of Tl at different pressures. As compression is increased, the population of electronic *d* (*s* and *p*) states around the Fermi level increases (decreases).

**Figure 1**

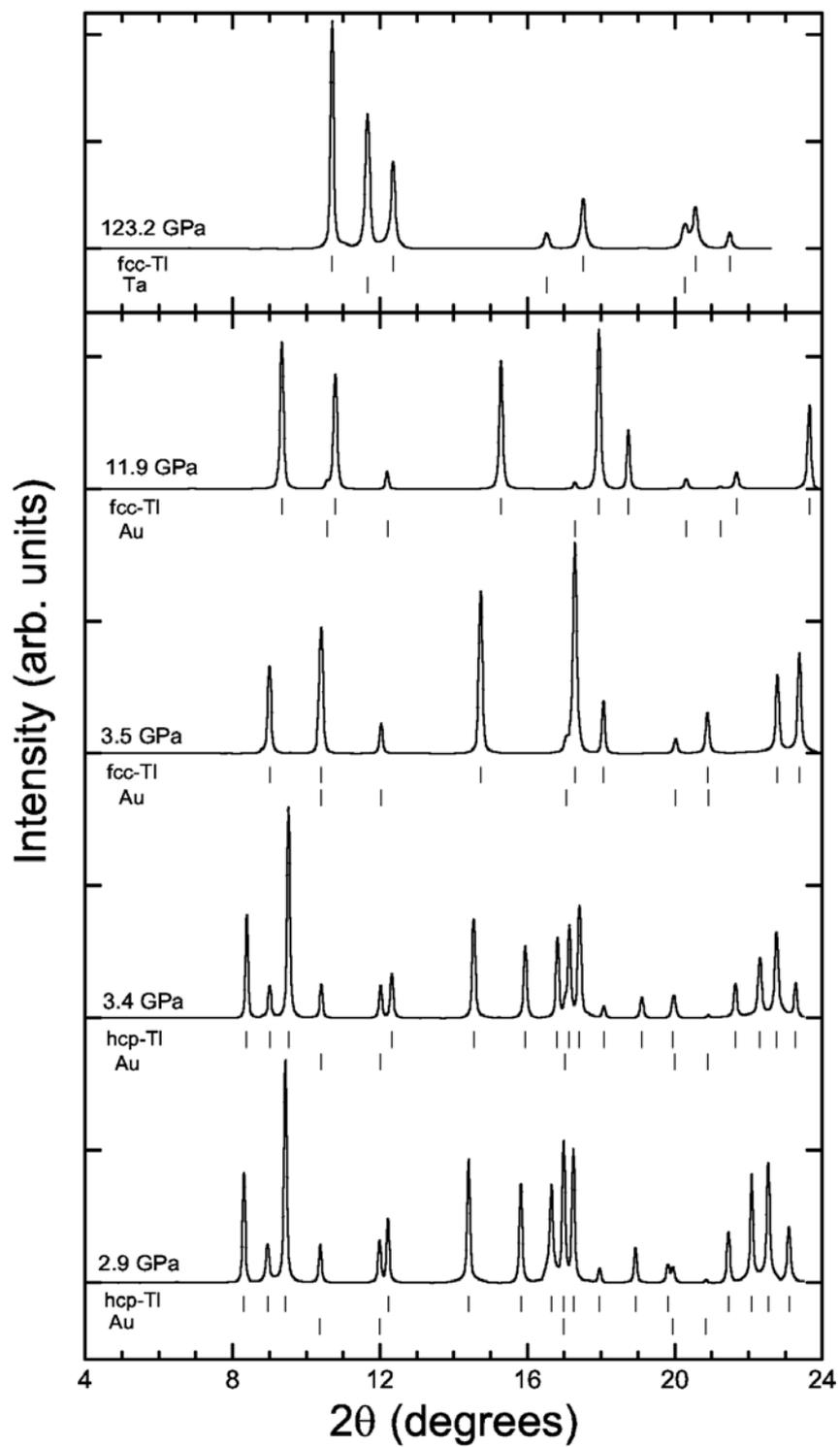

**Figure 2**

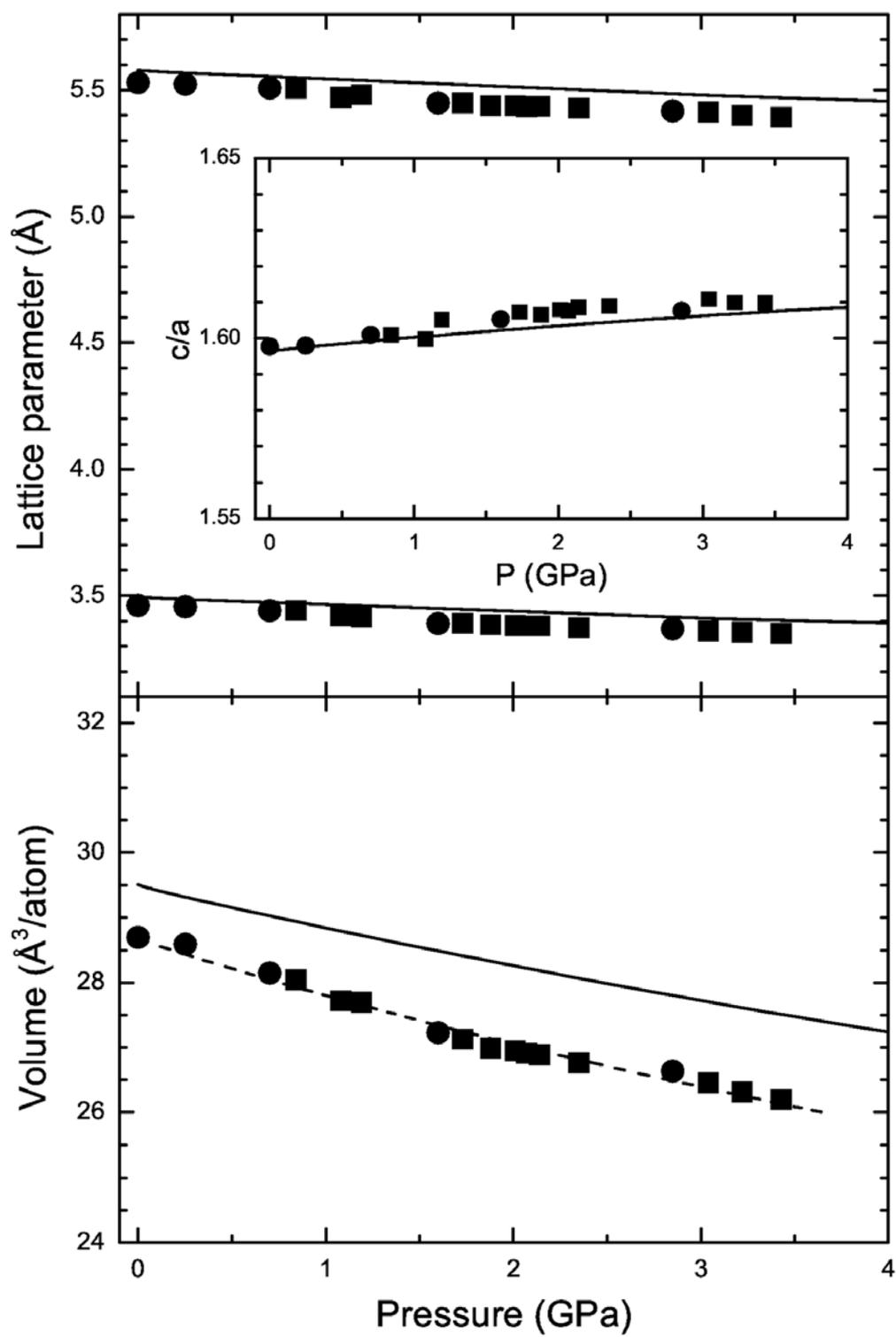

**Figure 3**

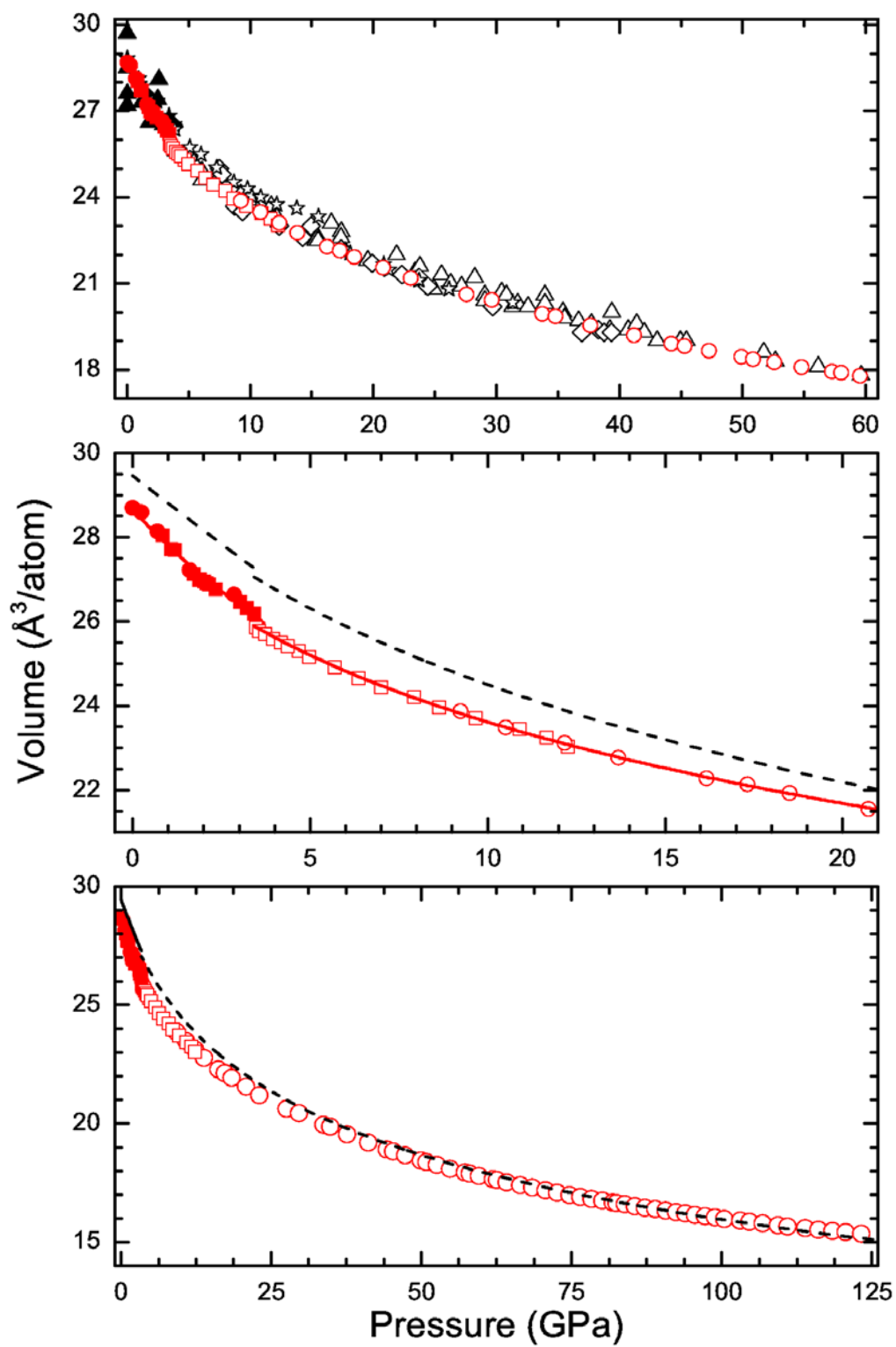

**Figure 4**

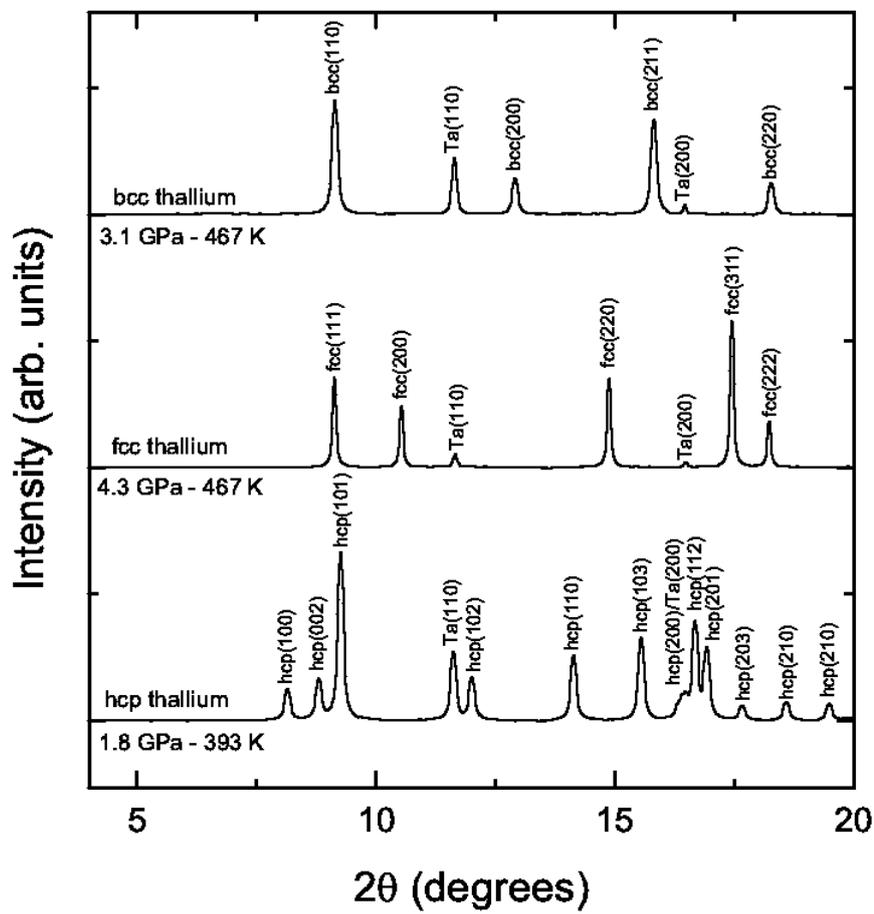

**Figure 5**

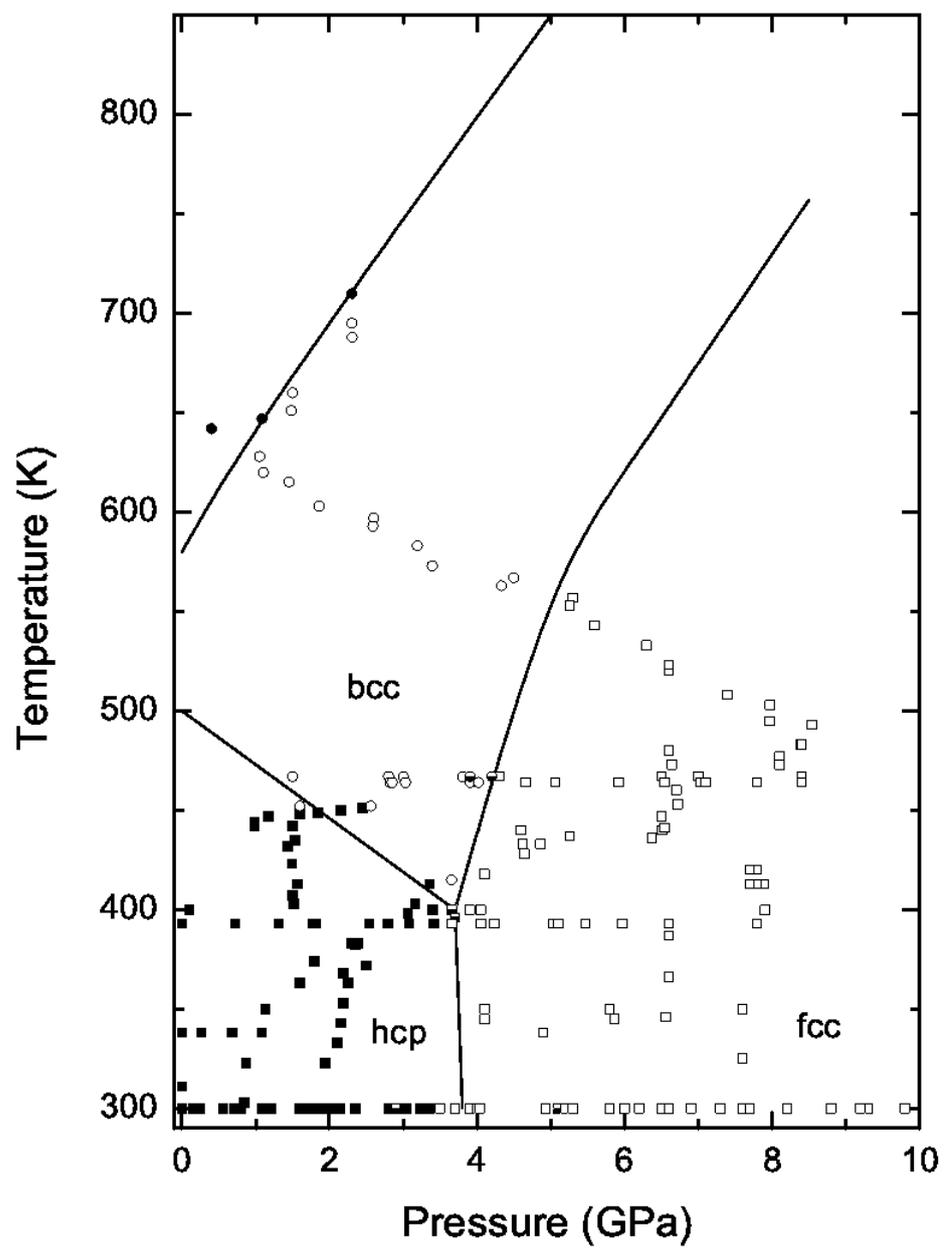

**Figure 6**

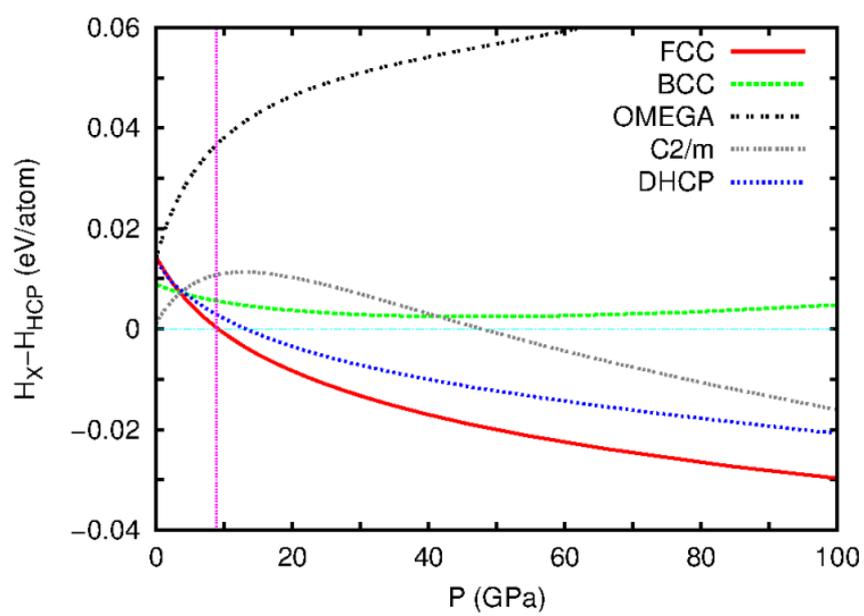

**Figure 7**

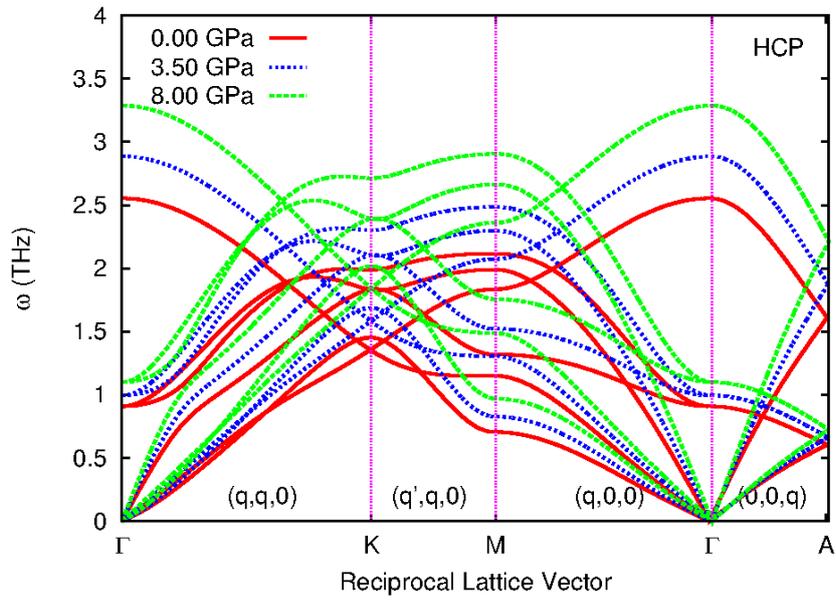

**Figure 8**

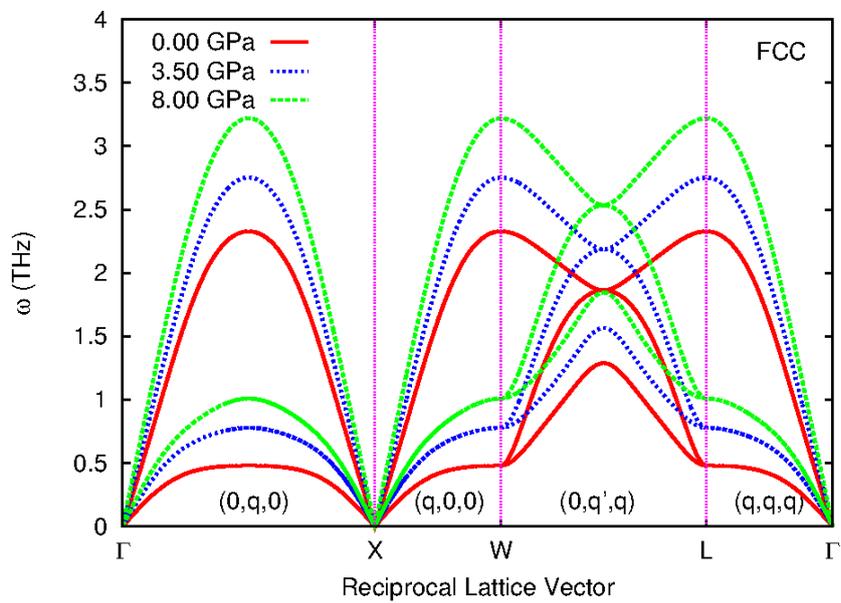

**Figure 9**

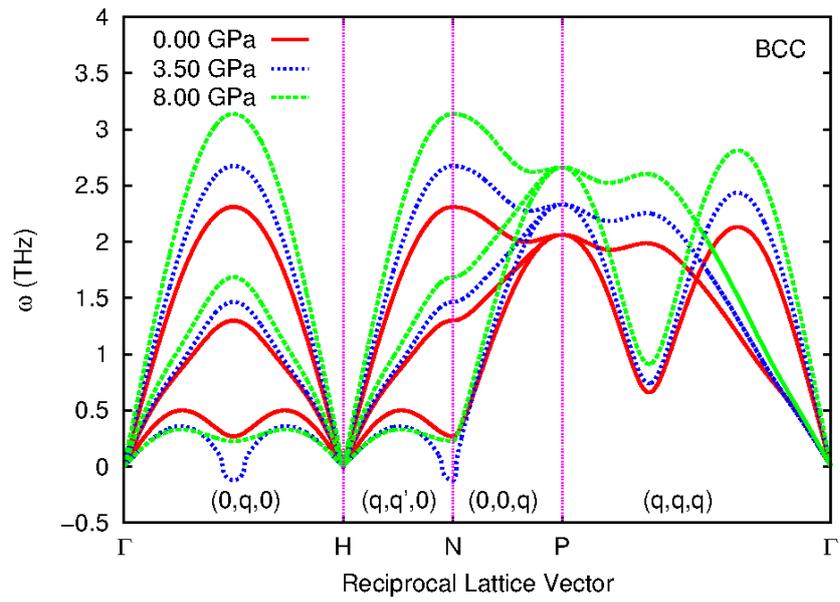

**Figure 10**

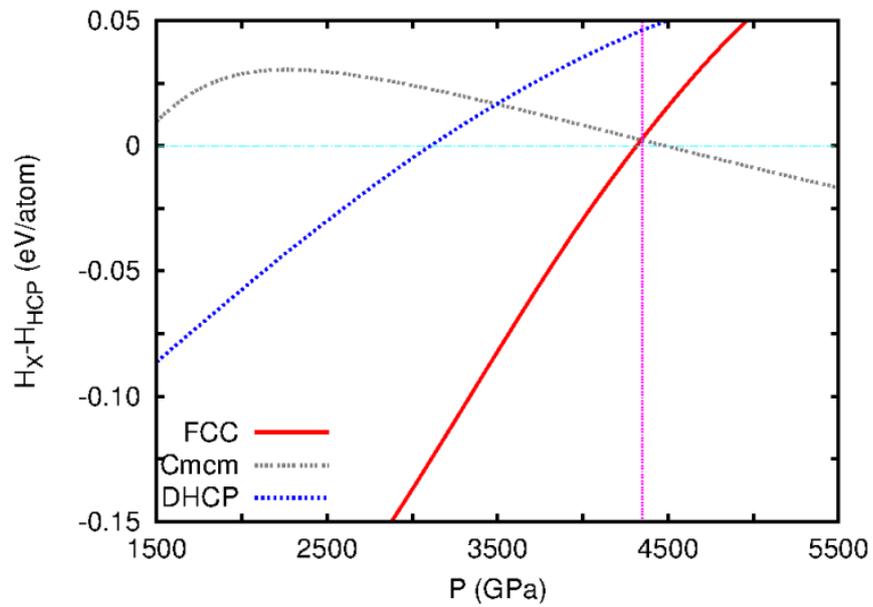

**Figure 11**

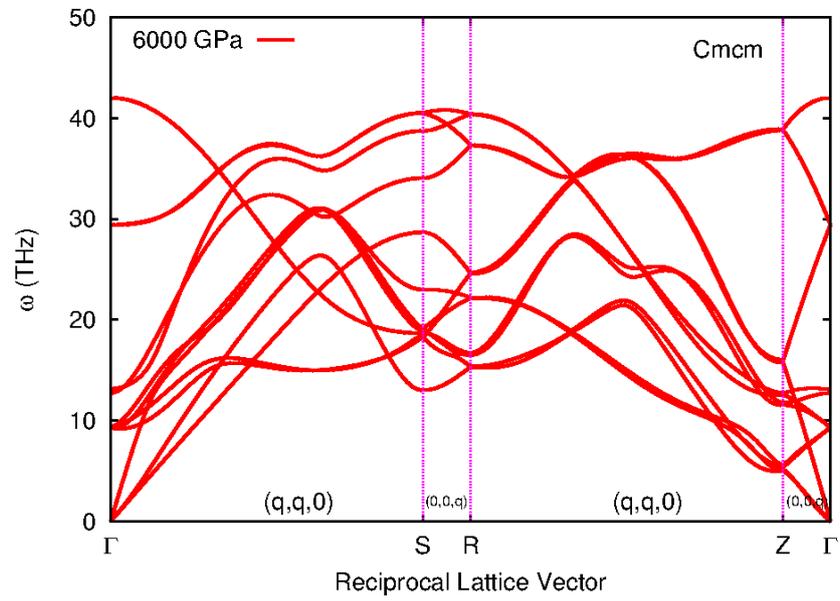

**Figure 12**

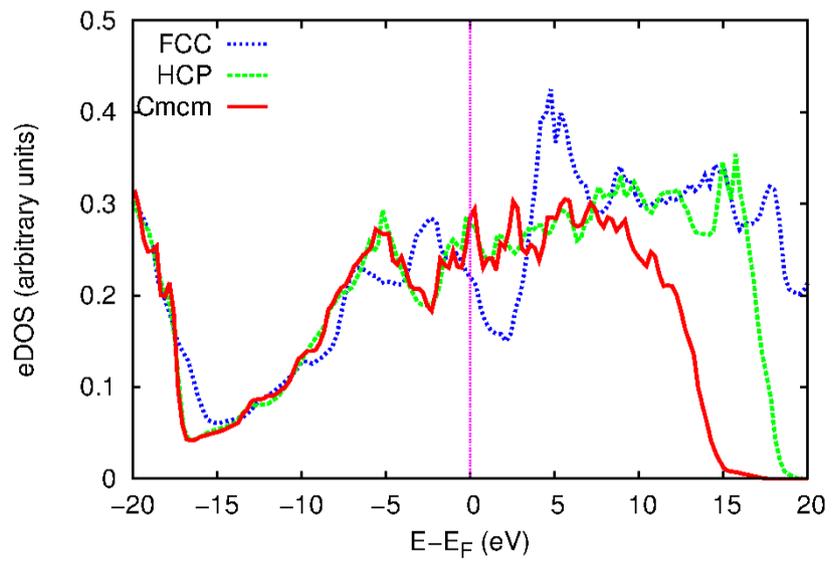

**Figure 13**

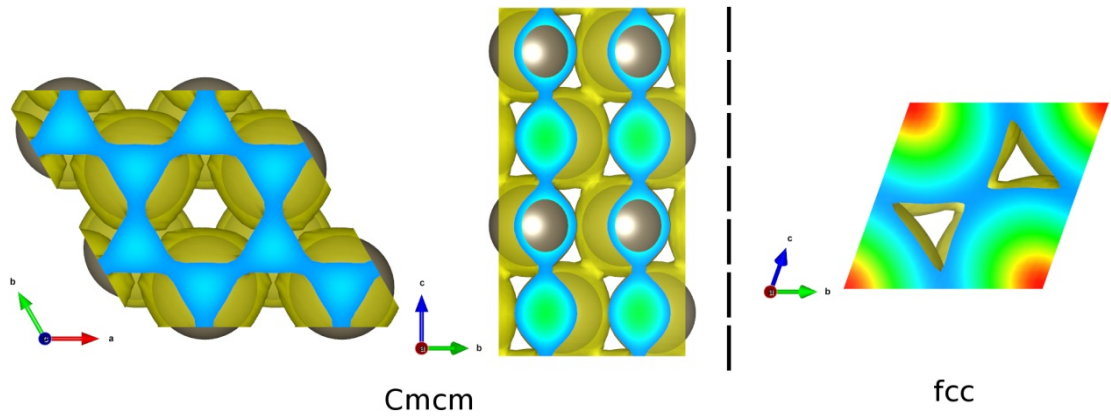

**Figure 14**

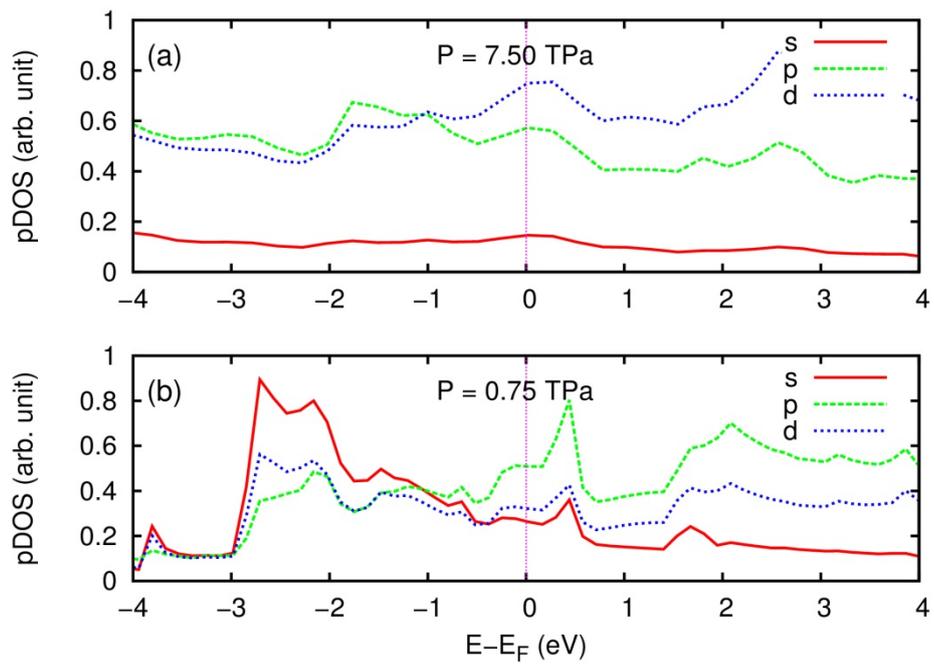